\documentclass[aps,prl,twocolumn,showpacs,superscriptaddress]{revtex4-1}
\usepackage{graphicx}
\usepackage{amssymb}
\usepackage{color}

\begin{document}

\title{Unconventional magnetostructural transition in CoCr$_2$O$_4$ at high magnetic fields}

\author{V. Tsurkan}

\affiliation{Experimental Physics V, Center for Electronic Correlations and Magnetism, Institute of Physics, University of Augsburg, D-86159, Augsburg, Germany}

\affiliation{Institute of Applied Physics, Academy of Sciences of Moldova, MD 2028, Chisinau, R. Moldova}

\author{S. Zherlitsyn}
\author{S. Yasin}
\affiliation{Hochfeld-Magnetlabor Dresden (HLD), Helmholtz-Zentrum Dresden-Rossendorf, D-01314 Dresden, Germany}

\author{V. Felea}

\affiliation{Institute of Applied Physics, Academy of Sciences of Moldova, MD 2028, Chisinau, R. Moldova}

\author{Y. Skourski}

\affiliation{Hochfeld-Magnetlabor Dresden (HLD), Helmholtz-Zentrum Dresden-Rossendorf, D-01314 Dresden, Germany}

\author{J. Deisenhofer}
\author{H.-A. Krug von Nidda}

\affiliation{Experimental Physics V, Center for Electronic Correlations and Magnetism, Institute of Physics, University of Augsburg, D-86159, Augsburg, Germany}

\author{J. Wosnitza}

\affiliation{Hochfeld-Magnetlabor Dresden (HLD), Helmholtz-Zentrum Dresden-Rossendorf, D-01314 Dresden, Germany}

\author{A. Loidl}

\affiliation{Experimental Physics V, Center for Electronic Correlations and Magnetism, Institute of Physics, University of Augsburg, D-86159, Augsburg, Germany}

\date{\today}

\begin{abstract}

The magnetic-field and temperature dependencies of ultrasound propagation and magnetization of single-crystalline CoCr$_2$O$_4$  have been studied in static and pulsed magnetic fields up to 14~T and 62~T, respectively. Distinct anomalies with significant changes in the sound velocity and attenuation are found in this spinel compound at the onset of long-range incommensurate spiral-spin order at $T_s$ = 27~K and at the transition from the incommensurate to the commensurate state at $T_l$~=~14~K, evidencing strong spin-lattice coupling. While the magnetization evolves gradually with field, step-like increments in the ultrasound  clearly signal a transition into a new magneto-structural state between 6.2 and 16.5 K and at high magnetic fields. We argue that this is a high-symmetry phase with only the longitudinal component of the magnetization being ordered, while the transverse helical component remains disordered. This phase is metastable in an extended \emph{H-T} phase space.

\end{abstract}

\pacs{62.65.+k, 72.55.+s, 75.50.Ee }

\maketitle

Multiferroic materials, which exhibit concomitant magnetic and ferroelectric order, are of great current interest, both from a fundamental as well as application-oriented view. They challenge our understanding of ordering phenomena, but in addition provide new functionalities in spintronics, since these dielectric and magnetic polarizations can be tuned either by external magnetic or electric fields \cite{Fiebig05,Chong07,Ramesh07,Tokura10}. Among multiferroics magnetic $AB_2X_4$ compounds with spinel structure attracted considerable interest revealing colossal magnetocapacitance and spontaneous dielectric polarization in the magnetically ordered state \cite{Hemberger05,Yamasaki06,Weber06,Murakawa08,Choi09}. The appearance of dielectric polarization is associated either with a non-collinear arrangement of spins, with charge order, or with magnetic ions moving off-center from their symmetric site positions in the lattice due to strong magneto-elastic effects.

Significant spin-lattice coupling and magnetic frustration are important features of spinels, specifically, of chromium oxides and chalcogenides. Despite quenched orbital moments these compounds reveal structural instabilities which are governed explicitly by ordering of spins, e.g., giant magnetostriction, negative thermal expansion \cite{Hemberger07}, and spin Jahn-Teller instabilities \cite{Lee00,Sushkov05,Hemberger06,Yamashita00,Tchernyshyov02}. In Cr spinels with only one magnetic sublattice, where the Cr$^{3+}$ ions are located solely on the pyrochlore lattice of the \emph{B} sites, it is well known that the oxides with dominating antiferromagnetic (AFM) exchange reveal strong geometrical frustration, while in the sulfides and selenides ferromagnetic (FM) exchange becomes important. ZnCr$_2$S$_4$ and ZnCr$_2$Se$_4$ are strongly frustrated due to competing AFM and FM interactions, with the ground state still being antiferromagnetic \cite{Rudolf07}. At low temperatures, the geometrically frustrated oxide spinels exhibit magnetization plateaus as function of an external magnetic field, with a 3-up 1-down spin configuration \cite{Ueda05,Ueda06,Matsuda07}. Theoretically, it has been suggested that these plateaus are stabilized by lattice distortions \cite{Penc04}. Indeed, in high-field ultrasonic experiments on CdCr$_2$O$_4$ these magnetization plateaus are also seen as broad strain plateaus \cite{Bhat11} indicating a stable structural configuration. In contrast, the high-field magnetization of bond-frustrated ZnCr$_2$S$_4$ and ZnCr$_2$Se$_4$ is rather smooth \cite{Tsur11, Felea12} but nevertheless these compounds show plateau-like anomalies in high-field ultrasound experiments.

In this Letter, we report new effects in the multiferroic spinel CoCr$_2$O$_4$ evidenced by anomalies in the ultrasound propagation in magnetic fields up to 62~T. CoCr$_2$O$_4$ crystallizes in the normal cubic $Fd\bar{3}m$ structure with Co$^{2+}$ (electronic configuration $3d^7$, spin $S = 3/2$)  and Cr$^{3+}$ ions ($3d^3$, $S = 3/2$) occupying tetrahedral and octahedral sites, respectively. A collinear ferrimagnetic spin order sets in below $T_C\approx$ 95~K,  followed by a transition into an incommensurate (IC) long-range conical-spiral state at $T_s~\approx$ 26 ~K \cite{Menyuk64,Tomiyasu04,Chang09}. Reports on the magnetic structure between $T_C$ and $T_s$ are controversial. Menyuk \textit{et al}. \cite{Menyuk64} and Tomiyasu \textit{et al.} \cite{Tomiyasu04} report on short-range order of the spiral state. In contrast, Chang \textit{et al.} \cite{Chang09} found a well-defined long-range ordered IC conical spin state. On further lowering the temperature an incommensurate-to-commensurate (C) lock-in transition occurs at $T_l\approx$ 14~K \cite{Funahashi87,Tomiyasu04,Chang09,Choi09}. The true spin configuration of this C spiral phase below $T_l$ remains to be clarified and is a matter of controversy \cite{Kaplan07}. A further remarkable property of CoCr$_2$O$_4$ is its multiferroicity. Coupling of charge and spin leads to anomalies of the dielectric constant at the magnetic
transitions \cite{Yamasaki06,Lawes06, Mufti10}. Importantly, a spontaneous dielectric polarization appears only in the IC spiral state below $T_s$ \cite{Yamasaki06} indicating a low-symmetry phase. Multiferroicity in CoCr$_2$O$_4$ was assigned to the inverse Dzyaloshinskii-Moriya (DM) interaction based on the fact that the observed modulation of the spins and the lattice have the same wave vector \cite{Arima07}.

From sound-velocity measurements in solids detailed information on the elastic constants can be derived. Besides structural phase transitions, ultrasound experiments probe e.g., spin-phonon coupling, orbital degrees of freedom or charge order \cite{Luthi05} and we observe corresponding anomalies at the transitions into the commensurate and incommensurate spiral spin states. Upon increasing the external magnetic field in the commensurate spiral ground state a clear transition into a new high-field phase is found. This phase remains metastable upon decreasing the magnetic field down to a lower critical field, below which the incommensurate-spiral ground state is recovered. We propose that this new high-field phase is characterized by a restored (possibly cubic) symmetry as a result of the external magnetic field overcoming the effects of the inverse DM interactions. In this scenario, only the longitudinal (collinear) component of the magnetization remains ordered, while the transverse component becomes disordered.

\begin{figure}[t]
\centering
\includegraphics[angle=0,width=0.45 \textwidth]{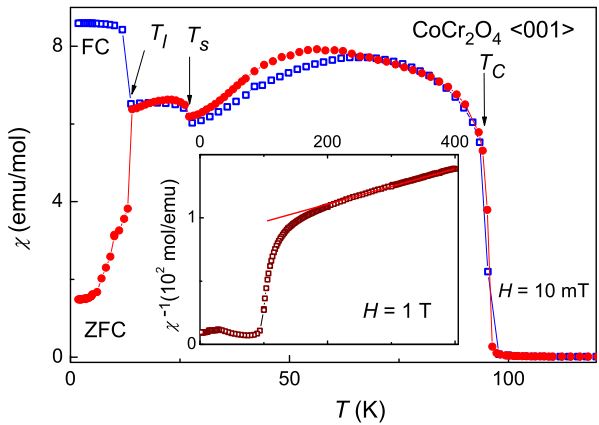}
\caption{(color online)  Temperature dependence of the zero-field-cooled (ZFC) (full circles) and field-cooled (FC) susceptibilities (open squares)  for CoCr$_2$O$_4$ measured in a field of 10~mT applied along the $\langle001\rangle$ axis. Inset: Temperature dependence of the inverse susceptibility measured in a field of 1~T. The solid line marks a Curie-Weiss fit at high temperatures. }
\end{figure}

High-quality CoCr$_2$O$_4$ single crystals were grown by chemical transport. Phase purity of the samples was checked by x-ray analysis. The elastic properties were studied by measurements of the velocity and attenuation of longitudinal waves with wave vector \textbf{\emph{k}} and polarization \textbf{\emph{u}} parallel to the $\langle001\rangle$ and $\langle111\rangle$ axes, which for a cubic crystal correspond to the elastic constants \emph{c}$_{11}$ and \emph{c}$_L$~=~1/3(\emph{c}$_{11}$+2\emph{c}$_{12}$+4\emph{c}$_{44}$), respectively. The measurements in static magnetic fields up to 14~T were performed for temperatures between 1.5 and 300~K. A phase-sensitive detection technique based on a pulse-echo method \cite{Wolf01} was used. Further ultrasound and magnetization measurements were performed in pulsed magnetic fields up to 62~T with rise time of 35~ms and pulse duration of 150~ms  in the temperature range from 1.5 to 20~K. The magnetic properties were also studied using a SQUID magnetometer (Quantum Design MPMS-5) in static fields up to 5~T.

In Fig.~1, the temperature dependencies of the zero-field-cooled, $\chi_{ZFC}$, and field-cooled, $\chi_{FC}$, susceptibilities for CoCr$_2$O$_4$ measured in a field of 10~mT applied along the $\langle001\rangle$ axis are shown. Both, $\chi_{ZFC}$ and $\chi_{FC}$ exhibit a step-like increase at the ferrimagnetic transition at $T_C\approx$ 95~K, and pronounced anomalies at the transition into the incommensurate spiral spin state at $T_s\approx$ 27~K and at the IC-to-C lock-in transition at $T_l\approx$ 14~K. Below $T_l$, a strong magnetization irreversibility and a history dependence appear due to the first-order magnetic phase transition at $T_l$ in agreement with neutron-diffraction data \cite{Chang09}. At temperatures above 250~K, the magnetic susceptibility follows a Curie-Weiss law with an asymptotic Curie-Weiss temperature $\theta_{CW} \approx$ -600~K. Deviations from the Curie-Weiss behavior emerge already below 250~K indicating the appearance of spin correlations (see inset of Fig.~1). The high ratio of $\mid\theta_{CW}\mid$/$T_C\approx 6$ points toward significant magnetic frustration due to competing ferromagnetic \emph{BB} (Cr-Cr) and antiferromagnetic \emph{AB} (Co-Cr) and \emph{AA} (Co-Co) exchange interactions that establish the spiral-spin configuration at low temperatures \cite{Lyons62,Ederer07}.

In Fig.~2, the sound velocity, \emph{v}, (right scale) and its relative change,  $\Delta \emph{v/v$_0$}$, (left scale) measured in different static magnetic fields for ultrasound waves propagating along the $\langle111\rangle$ axis are presented as function of temperature. In the high-temperature cubic phase and in zero magnetic field, \emph{v} $\sim$ 7.5~km/s, decreases with increasing fields and, contrary to what is observed in most solids, decreases towards lower temperatures. At $T_{C}$~=~95~K, \emph{v} exhibits a small anomaly only, indicative for a weak coupling of the longitudinal sound waves to the collinear spin structure which evolves below $T_{C}$. Between $T_{C}$ and the transition into the IC spiral spin state at $T_{s}$, the lattice softening is reflected by a considerable reduction of $\Delta \emph{v/v$_0$}$ of about 4\%. A similar strong softening was reported for the related frustrated spinel compounds CdCr$_2$O$_4$ and ZnCr$_2$S(Se)$_4$ \cite{Bhat11,Tsur11, Felea12} when entering the states with incommensurate spin configurations, which is accompanied by a structural transformation from cubic to tetragonal symmetry driven by a spin-Jahn-Teller effect \cite{Ueda05,Matsuda07,Yokaichiya09}. At $T_{s}$, \emph{v} shows a deep minimum followed by a step-like increase below $T_{s}$. On further decreasing temperature, \emph{v} exhibits a third anomaly at the transition into the commensurate state at $T_{l}$ with a hysteresis on cooling and warming (see inset in Fig.~2 for the elastic constant \emph{c}$_{11}$) confirming the first-order nature of this phase transition. With increasing magnetic field, the anomaly in the sound velocity at $T_{l}$ shifts to higher temperatures whereas the position of the anomaly at $T_{s}$ remains unchanged (inset of Fig.~2).

\begin{figure}[t]
\centering
\includegraphics[angle=0,width=0.45 \textwidth]{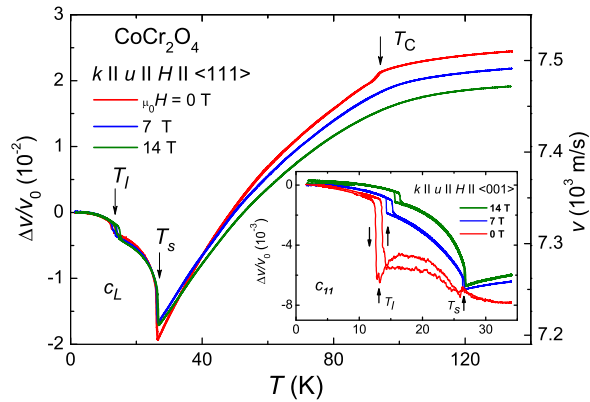}
\caption{(color online)  Temperature dependencies of the sound velocity, v, (right scale for zero field) and its relative change, $\Delta v/v_0$, (left scale)  for CoCr$_2$O$_4$ measured in different static magnetic fields for ultrasound waves propagating along the $\langle111\rangle$ axis.  The inset shows $\Delta v/v_0$ for waves propagating along the $\langle001\rangle$ axis at low temperatures for cooling and warming cycles. The arrows mark the magnetic phase transitions $T_{C}$, $T_{s}$ and $T_{l}$ observed at zero magnetic field.}
\end{figure}

\begin{figure}[t]
\includegraphics[angle=0,width=0.45 \textwidth]{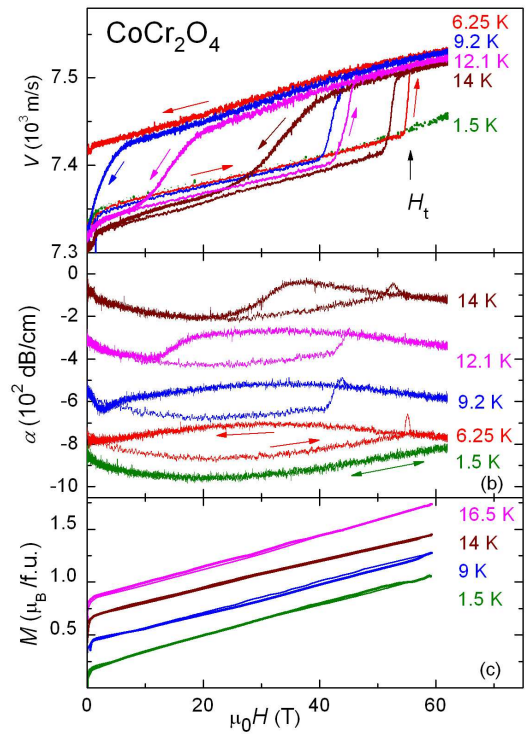}
\caption{(color online) (a) sound velocity, \emph{v}, (b) sound attenuation, $\Delta\alpha$, and (c) magnetization, \emph{M}, in CoCr$_2$O$_4$ at different temperatures vs. magnetic field applied along the $\langle111\rangle$ direction. Data for up and down field sweeps are shown. The vertical arrow marks the transition field $H_t$ at 6.25~K. The tilted arrows indicate the field sweep directions. The curves in panels (b) and (c) except that for 1.5~K are shifted along the vertical axis for clarity.}
\end{figure}

We now turn to the results obtained in high-magnetic fields presented in Fig.~3, where the sound velocity, \emph{v}, the sound attenuation, $\Delta\alpha$, and the magnetization, $M$, are shown as function of applied pulsed magnetic field for different temperatures. At the lowest temperature of 1.5~K, the sound velocity increases monotonously with magnetic field in a fully reversible manner. This reflects the field dependence of the sound velocity in the spiral spin configuration which is stable up to 62~T. At this temperature the sound attenuation changes nonmonotonously with field, decreasing up to 25~T, passing through a broad minimum, and increasing again towards higher fields without hysteresis. In contrast, at 6.25 K a step-like increase in the sound velocity and kink-like sharp maximum in the attenuation are found in a field $H_t$~=~55~T, signaling the transition into a new high-field phase with
significantly enhanced stiffness. We want to point out that the absolute value of the sound velocity in this state is close to \emph{v}($T_{C}$), the velocity in the vicinity of $T_C$ (see Fig. 2). This hints towards a similarity of the new high-field phase and the high-symmetry phase at higher-temperatures. Astonishingly, upon reducing the applied field the ground state is not recovered at the same magnetic-field values $H_t$, but the high-field phase remains metastable down to a lower transition field and seems to persist even to zero fields at 6.25~K. At 6.25~K, the field dependence of the sound attenuation on increasing fields up to $H_t$ resembles the behavior at 1.5 K. For fields above $H_t$  the slope of the attenuation changes sign, and for the down sweep a maximum appears close to 25~T. On further increasing temperatures, the transition field $H_t$ reaches a minimum close to 42~T at 9~K and increases again to 61~T at 16.5~K. The transitions at temperatures above 9.2 K are characterized by a large hysteresis which becomes narrower on increasing temperature.

Using our ultrasound results we constructed the low-temperature phase diagram of CoCr$_2$O$_4$ plotted in Fig.~4. The transition from the collinear ferrimagnetic (FiM) to the IC phase at $T_s$~ = ~27~K is almost field independent. The subsequent transition to the C spiral phase increases linearly with external magnetic field at least up to 14 T. For fields higher than 42~T a high-field phase appears. This high-field phase is metastable and prevails with decreasing field down to 0~T close to 6~K. At the lowest temperatures and at least up to 62~T  the commensurate spiral phase is the stable ground state.

The magnetization, \emph{M}, measured at several temperatures below $T_{s}$ shows an initial sharp ferromagnetic-like increase in fields below 2~T followed by a strictly linear dependence at higher fields. The ferromagnetic component has a value of 0.17~$\mu_B$/f.u. and changes slightly with increasing temperature from 1.5 to 27~K. Note that between 1.5 and 16.5~K, \emph{M} increases with an almost constant slope at high fields. Hence, the magnetization evidences no sign of a magnetic phase transitions, neither in passing the commensurate-to-incommensurate phase boundary nor in entering the high-field phase regime detected in the ultrasound experiments (Fig. 2, Fig. 3(a) and (b)). Similarly smooth magnetization curves have been reported for the multiferroic system Ni$_3$V$_2$O$_8$, when the external magnetic field is perpendicular to the magnetic easy axis \cite{Kenzelmann06,Wilson07}, while distinct magnetization steps are revealed parallel to the easy axis \cite{Wang11}. In CoCr$_2$O$_4$ where the magnetic anisotropy is by far smaller than in Ni$_3$V$_2$O$_8$ we even do not expect any magnetization steps for other orientations of the external field because the propagation of the spiral always follows the external field. Only nuclear magnetic resonance or neutron scattering at high fields could provide essential information about changes of the transverse magnetization component suggested in the proposed scenario.

This absence of concurrent features in the magnetization at first glance suggests that the high-field anomaly in the sound velocity is connected to purely structural changes. On the other hand, the step-like increase of the sound velocity to a value characteristic for the high-symmetry paramagnetic or collinear phase suggests an abrupt magnetic first-order transition from the low-symmetry spin spiral state to a high-symmetry possibly collinear spin state. In a rather narrow temperature regime this high-field phase remains metastable even in zero external field.

A similar field-induced effect with increase of the lattice stiffness was demonstrated recently for the frustrated ZnCr$_2$S(Se)$_4$ spinels and ascribed to the recovery of a high lattice symmetry \cite{Tsur11, Felea12}. However, in these systems a clear structural symmetry reduction is present in zero field. Also, in the frustrated oxide spinels, e.g., HgCr$_2$O$_4$ and CdCr$_2$O$_4$, the release of the geometrical frustration by magnetic field is accompanied by a structural transformation from tetragonal to cubic \cite{Ueda05,Matsuda07}.

Given the occurrence of a spontaneous polarization in CoCr$_2$O$_4$, we suppose that at low temperatures and zero magnetic field the lattice symmetry of CoCr$_2$O$_4$ must already be lower than cubic.

The external high-magnetic field will increase the Zeeman energy of the Cr$^{3+}$ ($g=2$) and Co$^{2+}$ ($g=2.23$) \cite{Abragam70,Tristan08} ions to about 7.3 meV and 8.1 meV, respectively, when using the minimal magnetic transition field of $H_t= 42$ T at 9 K. Theoretical estimates for the energy of the nearest-neighbor exchange interactions obtained from LSDA+U calculation are $J_{Cr-Co}= 3.6-4.3$ meV, $J_{Cr-Cr}=1.3-3.3$ meV, and $J_{Co-Co}=0.5-0.23$ meV (including the spin values)\cite{Ederer07}. Taking into account the number of interacting nearest neighbors, only the Cr-Co interaction is significantly larger than the magnetic field energy. On the other hand, the Co-Co interaction is easily overcome by the external field, while Cr-Cr interaction is comparable to the observed transition field.  Thus, the external magnetic field changes the balance between the competing exchange interactions and finally overrules the interaction responsible for the reduced symmetry.

In addition, the leading contribution of spin-lattice coupling and the origin of multiferroicity was assigned to arise from the inverse DM interaction and not from the conventional symmetric exchange striction \cite{Arima07}. The puzzling fact that we do not observe any anomaly in the field dependence of the magnetization $M(H)$, which is a measure of the longitudinal component of the magnetization, may become understandable by assuming that only the transverse component is affected in the high-field transition. With regard to the long debate on short-range and long-range magnetic order of the spiral component of the magnetization \cite{Tomiyasu04,Chang09}, we suggest that sufficiently strong external magnetic fields destroy the spiral order of the transverse magnetization component. This results in a quasi-paramagnetic state while the longitudinal component resides in collinear long-range order. Concomitantly to the breakdown of the transverse spiral order, the effects of the inverse DM interaction are canceled and the structure is released into presumably purely cubic symmetry. The high-field state remains metastable on lowering the field until the transverse magnetization is large enough to re-establish the spiral state. The neutron-scattering results \cite{Menyuk64,Tomiyasu04,Chang09} indicate that short-range and long-range ordered spirals are of similar ground-state energy providing the precondition for the metastability.

\begin{figure}[t]
\includegraphics[angle=0,width=0.45 \textwidth]{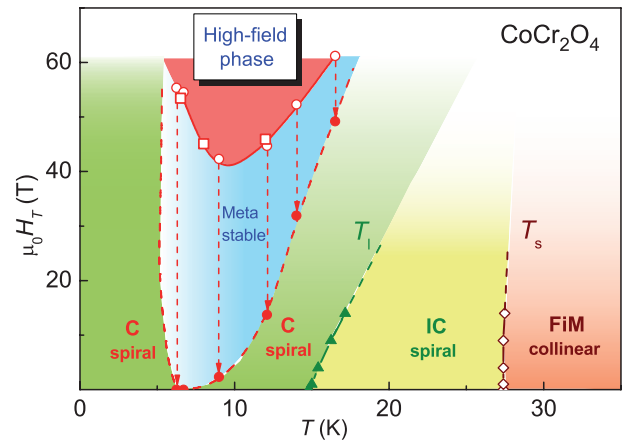}
\caption{(color online) Schematic low-temperature phase diagram of CoCr$_2$O$_4$. The phase boundaries between the metastable phase and helical C phase detected in the pulsed experiments are shown by dashed lines. The arrows connect the boundaries observed in increasing fields (open symbols) and decreasing fields (solid circles). Open squares and circles refer to the ultrasound data measured in pulsed fields applied along the $\langle001\rangle$ and $\langle111\rangle$ directions, respectively.}
\end{figure}

In conclusion, the present high-field studies of CoCr$_2$O$_4$ in magnetic fields up to 62~T reveal a remarkable phase diagram and unconventional magneto-elastic effects: While the transition from the paramagnetic into the collinear ferrimagnetic state reveals weak anomalies only, the lattice softens considerably on approaching the low-temperature spiral spin phase. The softening is continuous as a function of temperature but is abrupt as a function of magnetic field. The main result of our study is a fascinating island of stability of a field-induced phase at low temperatures with a minimal critical field of 42~T at 9.5~K. We speculate that this is a high-symmetry - probably cubic - phase, where the long-range order of the transverse component of the magnetization is suppressed  concomitantly with the exchange-driven structural distortions.

\begin{acknowledgements}

The authors thank Dana Vieweg for experimental support. This research has been supported by the DFG via TRR 80 (Augsburg - Munich) and by EuroMagNET II under the contract 228043.

\end{acknowledgements}

\end{document}